\documentclass[aps,prl,twocolumn,floatfix,superscriptaddress,amsmath,amssymb,longbibliography]{revtex4-2}
\usepackage{xcolor,bm,lineno,multirow,times}
\usepackage{physics}
\usepackage{rotating}
\usepackage{verbatim}
\usepackage{graphicx,tabularx}
\usepackage[sort&compress]{natbib}
\usepackage[T1]{fontenc}
\usepackage{soul}
\usepackage{dcolumn}
\usepackage{bm}
\usepackage[version=4]{mhchem}
\usepackage{booktabs}
\usepackage{gensymb}
\usepackage{color}
\usepackage[colorlinks=true,urlcolor=blue,linkcolor=magenta,citecolor=magenta]{hyperref}

\allowdisplaybreaks

\begin{document}
\title{Emulation of quantum correlations by classical dynamics in a spin-1/2 Heisenberg chain}
\author{Chaebin Kim}
\email{ckim706@gatech.edu}
\affiliation{School of Physics, Georgia Institute of Technology, Atlanta, Georgia 30332, USA}
\author{Martin Mourigal}%
\email{mourigal@gatech.edu}
\affiliation{School of Physics, Georgia Institute of Technology, Atlanta, Georgia 30332, USA}

\date{\today}

\begin{abstract}
We simulate the dynamical spin structure factor (DSSF) $\mathcal{S}({q},\omega)$ of the spin-1/2 Heisenberg antiferromagnetic chain using classical simulations. By employing Landau-Lifshitz Dynamics, we emulate quantum correlations through temperature-dependent corrections, including rescaling of magnetic dipoles and renormalization of exchange interactions. Our results closely match Quantum Monte-Carlo calculations for $k_{\rm B}T/J\!\gtrsim\!1$, extending the applicability of classical dynamics to the challenging case of gapless excitations. At higher temperatures, our simulations comply with general predictions for uncorrelated paramagnetic fluctuations in the infinite temperature limit. Entanglement witnesses derived from the quantum-equivalent DSSF act as sensitive diagnostics for the quantum-to-classical crossover. Their reliability stems from their dependence on spectral features alone, enabling classical dynamics to emulate quantum thresholds without genuine entanglement. This framework also reproduces transverse spin correlations in finite magnetic fields, in agreement with quantum simulations. Together, our results establish quantum-corrected classical dynamics as a scalable and predictive tool for interpreting scattering experiments and exploring quantum correlations in strongly correlated spin systems.
\end{abstract}
\maketitle


The spin-1/2 is Heisenberg antiferromagnetic chain is a central model in quantum magnetism, strongly correlated electron systems, and condensed matter physics in general~\cite{Giamarchi2003}. Its simplicity belies the rich and complex physics it embodies, making it an ideal model for exploring fundamental concepts such as deconfined fractional excitations, quantum criticality, and nontrivial entanglement structures. It serves as a theoretical testbed and proving ground for analytical and computational techniques and, perhaps surprisingly, remains of recurring experimental relevance, as it is realized in various quasi-one-dimensional magnetic materials. The ground state of this system can be understood exactly in the zero-temperature limit~\cite{Bethe1931}, a regime in which the excitations display a striking quantum phenomenon: the fractionalization of conventional magnons into pairs of spinons~\cite{Faddeev1981}. The presence of spinon pairs is associated with a continuum in both zero-temperature~\cite{Caux2006} and finite-temperature~\cite{Starykh1997,Grossjohann2009} response functions. 

In the last decade, continua of magnetic excitations have been evidenced in various magnetic materials by neutron and X-ray scattering experiments. This has led to rapid developments in numerical techniques to reproduce and explain the physical origin of these continua, which include spin fractionalization, magnon decay, cooperative paramagnetism, and chemical disorder. While quantum simulations offer high fidelity, they are computationally intensive and often limited in terms of system size, symmetry, and range of interactions. In contrast, classical dynamics does not rely on an accurate quantum representation of the ground state and is a surprisingly efficient tool for interpreting scattering experiments due to its speed and versatility in regimes where quantum simulations are impractical. However, the limits of applicability of classical dynamics are not yet fully understood.

In this Letter, we simulate the dynamical spin structure factor (DSSF) $\mathcal{S}({q},\omega)$ of the quantum (spin-1/2) Heisenberg antiferromagnetic chain (QHAC) using classical simulations~\cite{Keren1994,Huberman2008,Conlon2009,Taillefumier2014,Samarakoon2017,Zhang2019,Remund2022,Conlon2009,Remund2022}. We use Landau Lifshitz Dynamics (LLD), augmented with quantum-informed corrections~\cite{Dahlbom2022a,Dahlbom2022b}, to simulate the two-point spin correlations of this paradigmatic model and track their evolution in momentum-energy space as a function of temperature $T$ and uniform magnetic field $B$. We benchmark our results to existing quantum calculations previously obtained by Quantum Monte-Carlo (QMC)~\cite{Grossjohann2009}. Emulating such quantum calculations with classical dynamics relies on the correspondence principle between quantum and classical systems~\cite{Schofield1960,Huberman2008,Zhang2019,Dahlbom2024}. However, for a gapless system, this principle alone is not sufficient at finite temperatures, and we supplement it with two simple temperature-dependent corrections during thermalization and dynamical evolution: (i) a self-consistent rescaling of the length of the magnetic dipoles (so-called $\kappa(T)$ rescaling~\cite{Dahlbom2024}) ensures that the DSSF satisfies the correct spectral sum-rule; (ii) a modification of the antiferromagnetic exchange interaction $J$ (so-called $z(T)$ renormalization~\cite{Canali1992,Park2024}) ensures the first spectral moment of the DSSF matches with QMC calculations.

\begin{figure*}[t]
    \includegraphics[width=1.0\textwidth]{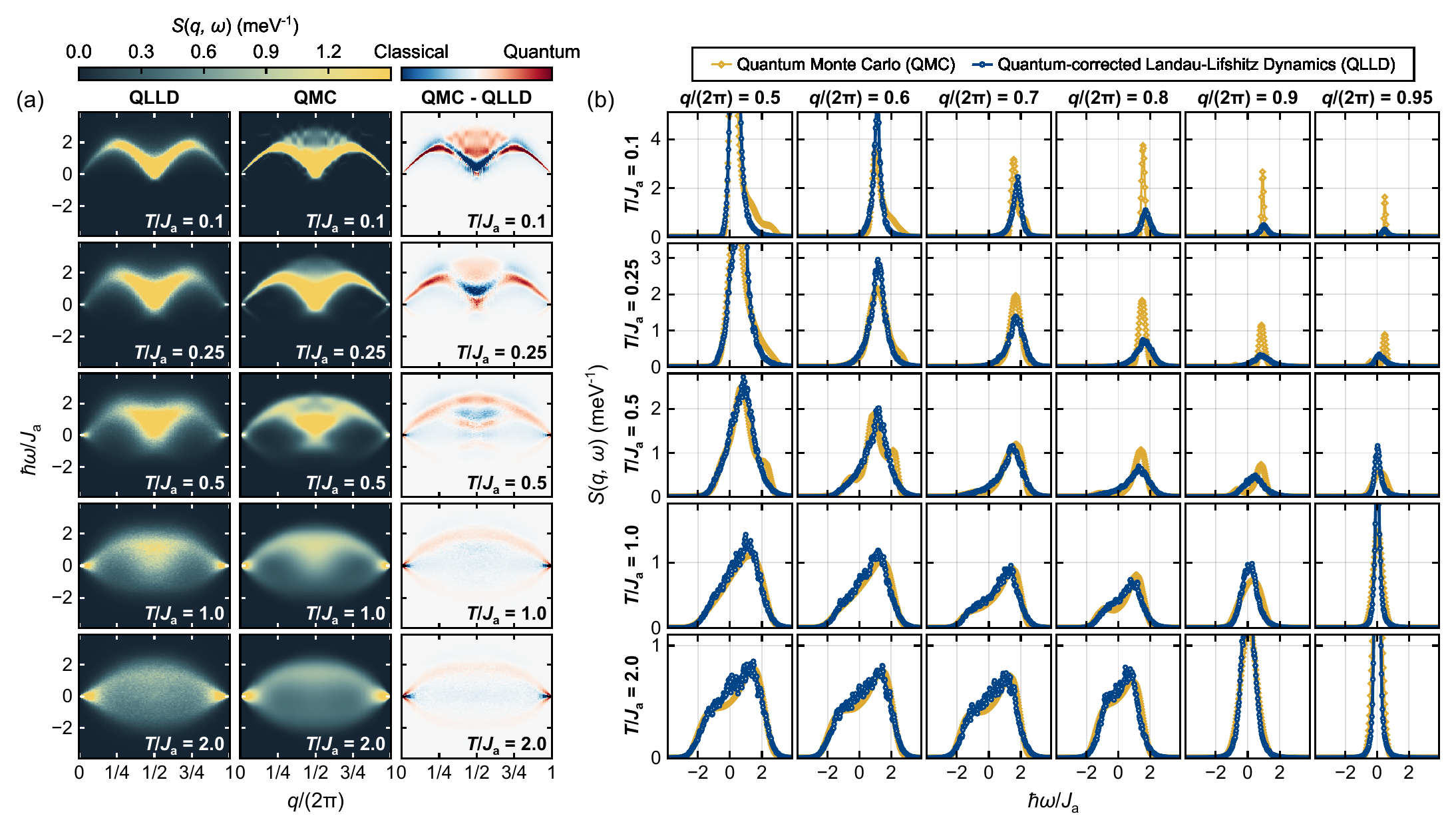}
    \caption{\label{fig:1} (a) Dynamical spin structure factor (DSSF) of the quantum Heisenberg antiferromagnetic chain (QHAC) as a function of temperature. The colorplot encodes the diagonal part of the DSSF $\sum_{\alpha} \mathcal{S}^{\alpha\alpha}(q,\omega)$ as a function of momentum and energy. The left column is obtained using quantum-corrected Landau-Lifshitz Dynamics (QLLD, this work), while the center column corresponds to Quantum Monte Carlo results from Ref.~\onlinecite{Grossjohann2009}. The right column is a plot of the difference between these two approaches. (b) A comparison between QLLD (blue lines) and QMC (gold lines) is made through the energy dependence of the DSSF at constant momenta, as indicated at the top of the column. The temperature is shown in each row. The dynamical structure factor is calculated in units of $J_a^{-1}={\rm meV}^{-1}$.}
\end{figure*}

Our results are fourfold. First, we obtain a quantum-equivalent DSSF that is overall identical to QMC results for $k_{\rm B}T/J\!\gtrsim\!1$; a regime where the continuum of magnetic excitations observed in neutron scattering measurements is emulated by the non-linear dynamics of the correlated paramagnet and approaches the universal spectral shape~\cite{deGennes1958} of Heisenberg paramagnets for $k_{\rm B}T/J\!\rightarrow\!\infty$. Second, at lower temperatures $k_{\rm B}T/J\!\lesssim\! 1$, our approach correctly captures the lower bound of the continuum using a large, and as expected \cite{desCloizeaux1962}, exchange renormalization. This result extends to transverse correlations in applied magnetic fields below saturation. Third, we identify our approach cannot correctly capture the lineshape of the high-energy spectral weight from spinon excitations at low temperature~\cite{Caux2006}, nor the sharpness of gapless incommensurate longitudinal spin correlations in applied field~\cite{Karbach1997,Stone2003}. Finally, we find that entanglement witnesses (EWs)~\cite{Hauke2016,Scheie2021,Laurell2021,Laurell2024} extracted from our DSSF closely mimic the quantum calculations at all temperatures. Given that our simulations rely on a product state without quantum coherence, entanglement, and non-locality, this merely reflects spectral mimicry and the limited discriminatory power of EWs when applied to the DSSF alone. Yet, values of EWs delineate the crossover between classical and quantum dynamics in our simulations and can thus be valuable in determining when a quantum treatment becomes necessary.

Our model Hamiltonian reads $\mathcal{\hat{H}} = J_a\sum_{\langle i,j \rangle} {\bf \hat{S}}_i \cdot {\bf \hat{S}}_j - g \mu_{\rm B} H \sum_i S^z_i$, where $J_a$ is the bare (or atomistic) antiferromagnetic exchange interaction between spin operators ${\bf \hat{S}}_i$. This model is approximately realized in a number of quasi-1D $S\!=\!1/2$ compounds including (but not limited to) KCuF$_3$~\cite{Tennant1993,Lake2005,Lake2013,Scheie2021}, copper pyrazine dinitrate (CuPzN)~\cite{Stone2003}, Sr$_{n}$CuO$_{n+1}$~\cite{Zaliznyak2004,Schlappa2012}, CuSO$_4\!\cdot\!5$D$_2$O~\cite{Mourigal2013}, and YbAlO$_3$~\cite{Wu2019,Kish2024}. These compounds span remarkably diverse chemical families and exchange interaction scales; as a result, a broad range of values for $k_{\rm B} T/J_a$ and $g\mu_{\rm B} H/J_a$ are accessible to detailed experimental investigations, especially using neutron scattering. High-fidelity theoretical predictions for the DSSF of this model have also been obtained using the Bethe Ansatz in the zero-temperature limit~\cite{Karbach1997,Caux2006}, as well as Quantum Monte-Carlo~\cite{Starykh1997,Grossjohann2009} and Matrix Product States techniques~\cite{Scheie2021,Laurell2021,Dupont2020,Menon2023} in various regimes. 

To calculate the DSSF with classical dynamics, we describe possible spin states as product states $| {\bf \Omega} \rangle = \otimes_i | {\bf \Omega}_i \rangle$, where $| {\bf \Omega}_i \rangle$ is a coherent state representing the $S\!=\!1/2$ dipole moment at site $i$ using a classical vector ${\bf \Omega}_i \equiv \langle {\bf \Omega}_i | \hat{\bf S}_i| {\bf \Omega}_i  \rangle$. The space-time dynamics of the classical dipoles are sampled from thermal equilibrium spin configurations evolving according to the Landau-Lifshitz equation of motion (LLD), as implemented in the {\scshape Sunny.jl}~\cite{Sunny} package~\cite{SI}. Because spin configurations are sampled with Boltzmann statistics, the classical DSSF must be corrected to emulate a quantum response at low temperatures. In the zero-temperature limit, the correction factor, $\hbar\omega / k_{\rm B}T \!\times\! [1-\exp(-\hbar\omega / k_{\rm B}T) ]^{-1}$, stems from the exact equivalence/correspondence between quantum and classical harmonic oscillators~\cite{Schofield1960,Huberman2008,Zhang2019,Remund2022,Dahlbom2024}. This correspondence is not exact at finite temperatures, given the non-linear nature of the classical dynamics. 

To progress, we implement two ad-hoc corrections based on the properties of spectral moments. First, the norm of each classical dipole is scaled to $|{\bf \Omega}_i(T)| = \kappa(T) S$, with a rescaling factor $\kappa(T)$ calculated self-consistently such that the corrected DSSF fulfills the zeroth-moment sum-rule $\sum_\alpha\int\!\int \mathcal{S}^{\alpha\alpha}({q},\omega) d\omega\,d{q} = S(S+1)$ at every simulation temperature~\cite{Dahlbom2024}. Second, the exchange interaction itself is renormalized by a factor $z(T)$~\cite{Canali1992,Park2024} to match the first-moment $\mathcal{K}(q) = \sum_\alpha\int\omega\mathcal{S}^{\alpha\alpha}(q,\omega)d\omega$ by QMC~\cite{Grossjohann2009}. We choose the first-moment  to match its well-known property in the paramagnetic regime $\sum_\alpha\int\omega\mathcal{S}^{\alpha\alpha}(q,\omega)d\omega = -(1/3N)\sum_{ij}J_{ij} \left< \mathbf{S}_i\cdot\mathbf{S}_j \right> [1-\cos{(q\!\cdot\! {r}_{ij})}]$, which directly reflects the strength of the exchange interactions \cite{Hohenberg1974,Stone2001}. This matching is achieved by a least-squares fitting between our DSSF calculations and QMC, with $z(T)$ as a fitting parameter [see Supplementary Information~\cite{SI} Sec.~I]. While these two corrections appear together in the classical Hamiltonian $\mathcal{H}_{cl}$ controlling the dynamics, effectively rescaling the exchange interaction as $J_a\!\rightarrow\!J_a z \kappa^2(T)$, only $\kappa(T)$ impacts the zeroth-moment sum-rule and thus both corrections play distinct roles, as we will show below. We refer to the resulting simulations as quantum-corrected Landau-Lifshitz dynamics (QLLD).

In Fig.~\ref{fig:1}, we compare the momentum- and energy-dependence of the DSSF calculated with QLLD to the QMC results from Ref.~\onlinecite{Grossjohann2009}. We focus on energies up to $\hbar\omega_{\rm max} = \pi J_a$ bounded by the upper branch of the two-spinon continuum~\cite{Karbach1997}. In the zero-temperature limit, this energy band is known to contain most of the spectral weight~\cite{Caux2006}. In the high-temperature limit, which we define here as $k_{\rm B}T/J_a\!\gtrsim\!1$, the dominant features of the DSSF are virtually identical between QLLD and QMC as shown on two-dimensional spectral plots [Fig.~\ref{fig:1}(a)-right/center] and constant-momentum line cuts [Fig.~\ref{fig:1}(b)]. The shape of the spectral continuum, which closely resembles recent experimental and numerical results on YbAlO$_3$~\cite{Kish2024} is accurately emulated by the classical dynamics of paramagnetic spins upon simple $\kappa(T)$ and $z(T)$ corrections [see SI~\cite{SI} Sec.~III]. We recall that the paramagnetic continuum is universal in the $k_{\rm B}T/J_a\!\rightarrow\!\infty$ limit and arises from the conservation of total magnetization for Heisenberg paramagnets~\cite{deGennes1958,Bunker1996}. Therefore we do not interpret it as evidence for coherent spinons~\cite{Krzysztof2025} or other fractional excitations. We nonetheless observe two small deviations between QLLD and QMC in this temperature regime, apparent in both line cuts [Fig.~\ref{fig:1}(b)] and subtraction spectral plots [Fig.~\ref{fig:1}(a)-left]. First, the drop in spectral intensity at higher energies appears sharper in QMC, which we hypothesize originates from details of the analytical continuation used in Ref.~\onlinecite{Grossjohann2009}. Second, in the vicinity of the zone center, the energy width of the DSSF at fixed momentum $q\rightarrow 0, 2\pi$ appears broader in QMC compared to QLLD; this is the only genuine deficiency of the QLLD approach above $k_{\rm B}T/J_a\!\gtrsim\!1$.

At lower temperatures, $k_{\rm B}T/J_a\!\lesssim\!1$, deviations between QLLD and QMC become more significant, as is apparent from difference spectral plots [Fig.~\ref{fig:1}(a)-right] and line cuts [Fig.~\ref{fig:1}(b)]. The strong signal at the lower boundary of the excitation continuum is well reproduced by QLLD, proving the effectiveness of our $z(T)$ renormalization approach. However, QLLD cannot explain the enhanced spectral weight at higher energies --- the hallmark of a multi-spinon continuum --- because of the classical nature of the underlying degrees of freedom and dynamics. This is particularly clear for $k_{\rm B}T/J_a\!=\!0.1$ and $k_{\rm B}T/J_a\!=\!0.25$, from the broad red area between $0.4 \leq q/(2\pi) \leq 0.6$ in Fig.~\ref{fig:1}(a)-right which signals the failure of QLLD to produce an asymmetric energy continuum. Second, QLLD produces a spectral distribution considerably broader than QMC near the Brillouin zone center. This effect is particularly striking in the line cuts of Fig.~\ref{fig:1}(b), which shows that the momentum-dependent peak broadening and suppression in QLLD is much more pronounced than that of QMC. We ascribe this discrepancy to the incapacity of QLLD to capture the coherent propagation of spinon excitations pointed out in Ref.~\onlinecite{Kish2024}. 

\begin{figure}[t]
\includegraphics[width=1.0\columnwidth]{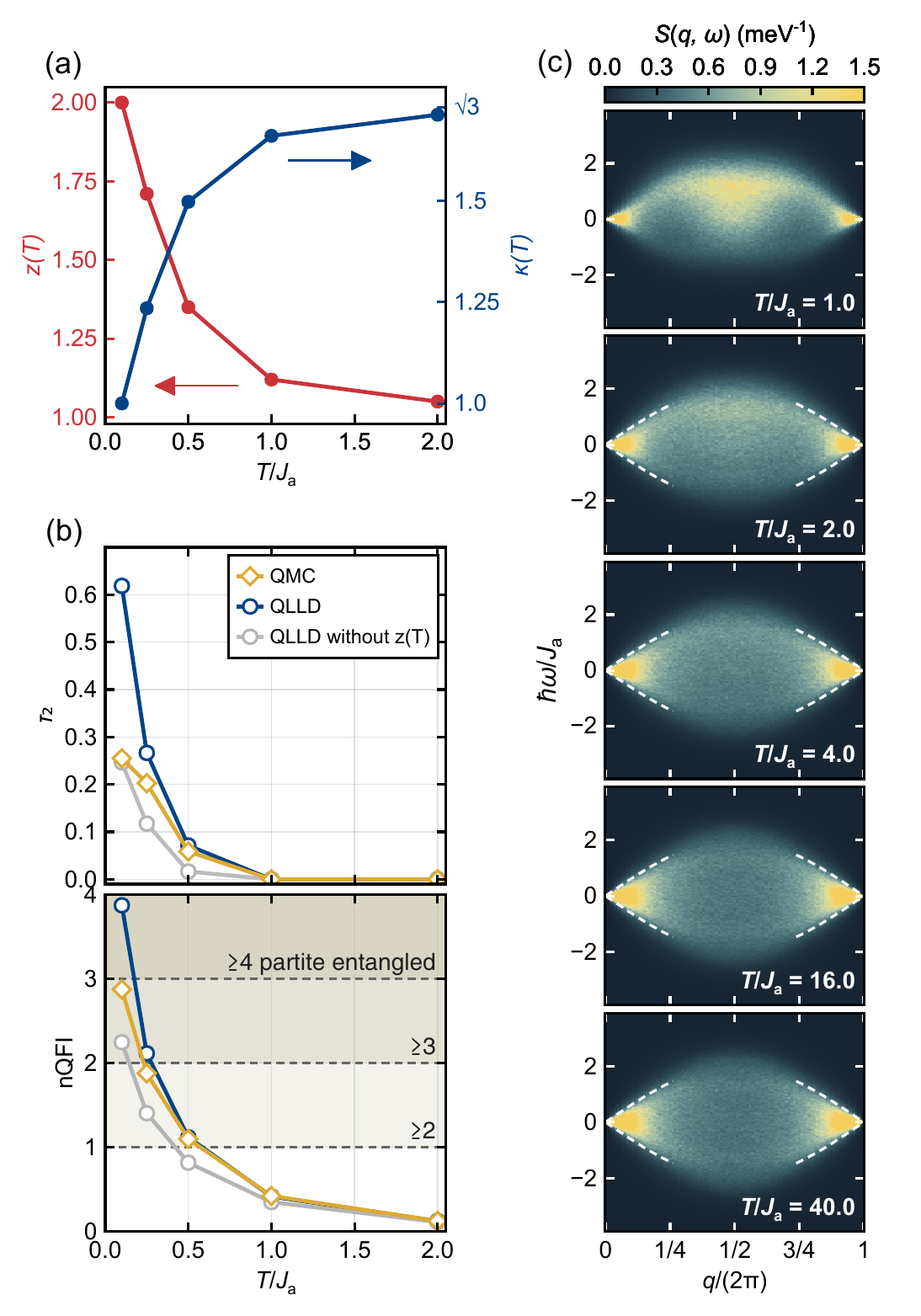}
    \caption{\label{fig:2} (a) Temperature dependence of the exchange renormalization factor $z(T)$ (red line) and moment rescaling factor $\kappa(T)$ (blue line) extracted from QLLD simulations with $z(T)$ fitted from QMC results. The y-scale corresponding to each quantity is highlighted with the corresponding colors and arrows. (b) Temperature dependence of entanglement witnesses (two-tangle and Quantum Fisher Information) extracted from QMC (gold lines) and LLD simulations, the latter with  (blue lines) and without (gray lines) $z(T)$-renormalization. The upper panel shows the temperature dependence of the two-tangle and the bottom panel shows the temperature dependence of the normalized quantum Fisher information. Grey dashed lines indicate the integer number of normalized QFI, which guarantees the lower bound of the $(n+1)$ multipartite entanglement. (c) High-temperature simulations of the DSSF using QLLD. For temperatures higher than $k_{\rm B}T/J_a\!=\!2$, we used $z(T)\!=\!1$ and $\kappa(T)\!=\!\sqrt{3}$ from Fig.~\ref{fig:2}(a). White dashed lines indicate the boundaries of the universal continuum for high-temperature Heisenberg paramagnets $\sqrt{2\left(1-\cos{q}\right)}$~\cite{deGennes1958}.}
\end{figure}

\begin{figure*}[t!]
\includegraphics[width=1.0\textwidth]{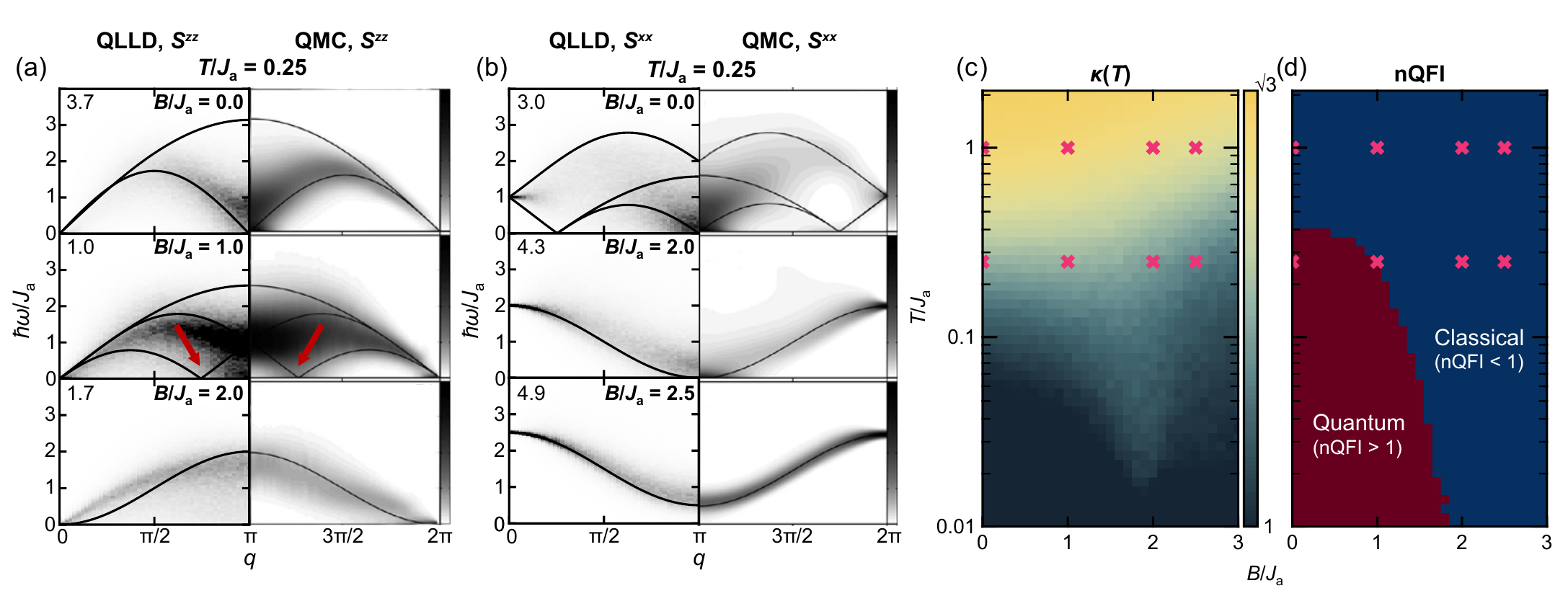}
    \caption{\label{fig:3} Field dependence of DSSF by QMC and QLLD at $k_{\rm B}T/J_a\!=\!0.25$. (a),(b) Longitudinal DSSF ${\cal S}^{zz}(q,\omega)$ and Transverse DSSF ${\cal S}^{xx}(q,\omega)$ of QLLD and QMC method, respectively. For the $B/J_a < 2$, black lines indicate the excitation boundary from the Muller ansatz at zero temperature. For the critical field $B/J_a \geq 2$, dispersion becomes $\omega(q) = 1-cos(q)$ for longitudinal DSSF and $\omega(q) = B/(2J_a)+cos(q)$ for trnasverse DSSF . The temperature is shown at the top of the column, and the magnetic field is given in each box. The red arrows in (a) indicate the expected incommensurate correlation from the QMC and Muller ansatz. The QMC results are from \cite{Grossjohann2009}. (c),(d) Temperature-field dependence of spin-length rescaling factor $\kappa(T)$ and nQFI. Magenta cross markers indicate the temperature-field condition of QMC and QLLD results. Red colored area in (d) indicates the nQFI $\geq$ 1 and blue area shows nQFI < 1.}
\end{figure*}

We now turn to details of our exchange renormalization procedure, which yields a temperature-dependent exchange interaction $J(T) = z(T)J_a$ reflecting a genuine quantum effect not captured by the classical dynamics and necessary to match the first moment $\mathcal{K}(q)$. Since the value of the first moment is related to the exchange interactions, its temperature-dependence can be used to determined the renormalization of the exchange interaction. In Fig.~\ref{fig:2}(a), we empirically compare the temperature dependence of $z(T)$ and $\kappa(T)$. Both quantities display an exponential-like behavior. The moment rescaling $\kappa(T)$ increases with temperature from $\kappa(T\rightarrow 0) = 1$ to $\kappa(T\rightarrow \infty) = \sqrt{S(S+1)}/\sqrt{S^2}=\sqrt{3}$ with an activation function that appears universal across different models [see SI~\cite{SI} Sec.~I]. The exchange renormalization $z(T)$ appears anti-correlated with this behavior and increases with decreasing temperature from $z(T)\!=\!1.03$ for $k_{\rm B}T/J_a\!=\!2$ to $z(T)\!=\!2.0$ for $k_{\rm B}T/J_a\!=\!0.1$. This behavior is consistent with general expectations but deviates from naive expectations of $z(T\rightarrow0)\!=\!\pi/2$, consistent with Des Cloizeaux-Person~\cite{desCloizeaux1962,Mourigal2013} and $z(T\rightarrow\infty)\!=\!1$. The $z(T)$ correction is essential for gapless systems, where spectral redistribution cannot be captured by $\kappa(T)$ alone, unlike in previous studies on gapped magnets~\cite{Dahlbom2024,Dahlbom2024b}. Fortunately, $z(T\rightarrow0)$ can generally be derived from perturbative expansions such as $1/S$-corrected spin-wave theory~\cite{Canali1992} or by effective field theory approaches \cite{Chakravarty1989}.

We now extend our examination to entanglement witnesses~\cite{Hauke2016,Scheie2021,Laurell2021,Laurell2024} which have recently been proposed as a general protocol to certify the presence of bipartite and multipartite quantum entanglement in spin systems through the momentum- and energy-dependence of the DSSF. These approaches have been discussed extensively for the QHAC in the context of experimental data and numerical tensor network results~\cite{Scheie2021,Laurell2021,Dupont2020,Menon2023,Kish2024}. Here, we test two EWs: the two-tangle ($\tau_2$) characterizing the total entanglement carried by pairwise correlations, and the normalized quantum Fisher information (nQFI), which can be interpreted as the lower bound of the multipartite entanglement [See SI~\cite{SI} Sec.~II]. Fig.~\ref{fig:2}(b) shows $\tau_2(T)$ and nQFI$(T)$ extracted from three different numerical datasets: QMC, QLLD, and QLLD without $z(T)$ renormalization. The overall trend and values for these EWs are similar across all three approaches and resemble previous results. 

Surprisingly, the DSSF calculated using classical dynamics --- even without exchange renormalization --- can produce above threshold values for $\tau_2$ and nQFI for $k_{\rm B}T/J_a\!<\!1$. This does not signify genuine entanglement in classical simulations. Instead, it highlights the limited ability of entanglement witnesses based on spectral features to distinguish truly quantum correlations when applied to the DSSF. We postulate that the resemblance between classical and quantum coupled harmonic oscillators allows classical dynamics to emulate the values of these witnesses and give the appearance of quantum entanglement~\cite{Weedbrook2012}. Nonetheless, we remark that the temperature regime for which significant deviations emerge between QMC and QLLD [$k_{\rm B}T/J_a\!<\!1$, Fig.~\ref{fig:1}(a)] coincides with thresholds being exceeded for both pair and multipartite entanglement [Fig.~\ref{fig:2}(b)]. This observation suggests that entanglement witnesses can serve as a self-consistent diagnostic tool for the validity of classical emulation of the DSSF.

We now turn to the high-temperature regime $k_{\rm B}T/J_a\!>\!2$, for which we have extended QLLD simulations up to $k_{\rm B}T/J_a\!=\!40$ [Fig.~\ref{fig:2}(c)]. Classical dynamics accurately represent the DSSF in the $T\rightarrow\infty$ limit. The pinched-oval gapless spectral continuum observed for $k_{\rm B}T/J_a\!=\!4$, $16$, and $40$, is thus the hallmark of uncorrelated paramagnetic fluctuations. As was pointed out by De Gennes in 1958~\cite{deGennes1958}, this behavior can be understood from general principles and is readily captured by classical dynamics for other Heisenberg spin systems~\cite{Bunker1996}. For the QHAC in the limit of $q\!\rightarrow\!0$, the Fourier transformed spin operator $\hat{S}^z_q = 1/\sqrt{L}\sum_j \exp(i q r_j) \hat{S}^z_j$ commutes with the Hamiltonian at any temperature, $[\hat{S}^z_{q=0}, \mathcal{H}]_T=0$, as the total magnetization $\hat{M}^z_{\rm tot} = \sqrt{L}\hat{S}^z_{q=0}$ is conserved. Thus, the DSSF $\mathcal{S}^{zz}(q=0,\omega,T) = \langle[\hat{M}^z]^2\rangle/L\,\delta(\omega)$ is purely elastic in the $q\!\rightarrow\!0$ limit. This argument can be extended from the Brillouin zone center by considering the spectral weight distribution's second moment (or variance), $\sigma^2(q,T)$. Clearly, as discussed above, $\sigma(q\!\rightarrow\!0,T)\!=\!0$ for any temperature. For finite momentum, the variance can only be approximated in the infinite temperature limit by considering the system's translational, inversion, and rotational invariance. This yields $\sigma^2(q,T\!\rightarrow\!\infty)=2J_a (1 - \cos q)$~\cite{deGennes1958} for the QHAC. This readily explains the spectral distribution (which we propose to call the ``De Gennes continuum'') observed in our QLLD simulations and recent experiments on YbAlO$_3$~\cite{Kish2024}. This is a universal feature of diffusive precessional spin dynamics~\cite{Conlon2009,Grossjohann2010} in an uncorrelated Heisenberg paramagnet.

We finish by simulating the transverse and longitudinal contributions to the DSSF at finite temperature and finite magnetic field and compare our results with QMC~\cite{Grossjohann2009}. Results in applied magnetic fields of $B/J_a\!=\!g\mu_{\rm B}H/J_a\!=\!1.0, 2.0, 2.5$ (the saturation field of the quantum model is $B_s/J_a=2.0$) are shown in Fig.\ref{fig:3} for $k_{\rm B}T/J_a\!=\!0.25$ [See SI~\cite{SI} Sec.~IV for the $k_{\rm B}T/J_a\!=\!1$ case]. In the low-temperature limit, it is well known that incommensurate magnetic correlations develop even for realistic systems~\cite{Stone2003,Wu2019}. This phenomenon can be pictured in the language of fermionic spinons as originating from a shift in chemical potential away from half-filling, yielding low-energy excitations incommensurate with the crystallographic cell. As expected, the incommensurate longitudinal fluctuations cannot be captured by classical dynamics, see the red arrows for $B/J_a\!=\!1$ at $k_{\rm B}T/J_a\!=\!0.25$ in Fig.\ref{fig:3}(a).  However, surprisingly, the transverse correlations simulated by QLLD look overall similar to QMC for all fields, even when $k_{\rm B}T/J_a < 1$.

To understand this agreement, we mapped the spin-length rescaling factor $\kappa(T)$ and the nQFI as functions of temperature and magnetic field [see Fig.~\ref{fig:3}(c),(d)]. Both metrics clearly reveal the saturation field. In the low-field regime, $\kappa(T)\approx1$ at low temperatures, but begins to deviate as the field approaches saturation. Similarly, nQFI mirrors a phase diagram, identifying the crossover from classical to quantum behavior at nQFI $=1$ and highlighting regions where quantum dynamics dominate (nQFI $\geq 1$). Notably, nQFI also locates the saturation field across the entire temperature range. This mapping explains the close agreement between QLLD and QMC at $B/J_a = 1$ and $k_{\rm B}T/J_a = 0.25$ where nQFI $\approx1$.Crucially, the diagnostic power of $\kappa(T)$ and nQFI requires no exchange renormalization, making them useful and universal indicators for the validity of classical simulations.

Our work reinforces the emerging consensus that classical LLD, when corrected by temperature-dependent spin rescaling and exchange renormalization, can faithfully emulate the DSSF of quantum spin systems at finite but small temperatures~\cite{Samarakoon2018, Steinhardt2021, Knolle2022, Park2024}. It is remarkable that this approach works down to $k_{\rm B}T/J_a\!\approx\!1$ in the extreme case of the gapless correlations of the quantum Heisenberg antiferromagnetic chain. Our approach reproduces expected behavior above $k_{\rm B}T/J_a\!\approx\!1$, including the universal De Gennes continuum at high temperatures, and captures transverse correlations even in a finite field. While classical dynamics cannot reproduce incommensurate quantum coherence or multi-spinon continua, entanglement witnesses derived from the DSSF serve as predictive diagnostics for identifying the breakdown of classical emulation. Crucially, these diagnostics require no exchange renormalization, making them broadly applicable. Our results establish quantum-corrected classical dynamics as a practical and scalable framework for interpreting scattering experiments, offering a complementary method to diffuse scattering~\cite{Paddison2012}, dynamical high-temperature expansions~\cite{burkard2025dynamiccorrelationsfrustratedquantum}, and saturated field measurements~\cite{Mourigal2013} for systems where the saturation field is too high or reached only asymptotically. More broadly, this approach provides a systematic way to probe the quantum-to-classical crossover in strongly correlated spin systems.

\begin{acknowledgements}
This work was supported by the US Department of Energy, Office of Science, Basic Energy Sciences, Materials Sciences and Engineering Division under award DE-SC-0018660. The authors are especially grateful to Prof. Wolfram Brenig for sharing their QMC data and are indebted to Prof. Cristian Batista for his numerous insights. The authors also thank Dr. David Dalhbom and Pyeongjae Park for helpful discussions about $\kappa(T)$ rescaling. 
\end{acknowledgements}

Data availability: All data used to create the figures in this article are available in \cite{data}

\bibliography{refs-q2c1d-final}

\end{document}